\documentclass[twocolumn,12pt]{article}

\usepackage[authoryear]{natbib}
\usepackage{amssymb}
\usepackage{graphicx}

\begin{document}

\title{Quantitative analysis of spirality in elliptical galaxies}

\author{Levente Dojcsak, Lior Shamir\footnote{lshamir@mtu.edu} \\ Lawrence Technological University, Southfield, Michigan}

\maketitle

\begin{abstract}
We use an automated galaxy morphology analysis method to quantitatively measure the spirality of galaxies classified manually as elliptical. The data set used for the analysis consists of 60,518 galaxy images with  redshift obtained by the Sloan Digital Sky Survey (SDSS) and classified manually by {\it Galaxy Zoo}, as well as the RC3 and NA10 catalogues.  We measure the spirality of the galaxies by using the Ganalyzer method, which transforms the galaxy image to its radial intensity plot to detect  galaxy spirality that is in many cases difficult to notice by manual observation of the raw galaxy image. Experimental results using manually classified elliptical and S0 galaxies with redshift $<$0.3 suggest that galaxies classified manually as elliptical and S0 exhibit a nonzero signal for the spirality. These results suggest that the human eye observing the raw galaxy image might not always be the most effective way of detecting spirality and curves in the arms of galaxies.
\end{abstract}

{\bf keywords}:
Galaxies: elliptical and lenticular -- Techniques: image processing.

\section{Introduction}
\label{introduction}

Galaxy morphology is studied for the purpose of classification and analysis of the physical structures exhibited by galaxies in wide redshift ranges in order to get a better understanding of the structure and development of galaxies. While significant research has been done to study the morphology of galaxies with spiral arms \citep{Loveday96, Bal08b,Nair2009,Nair2010}, research efforts have been focused also on the analysis of elliptical and S0 galaxies using photometric measurement of the electromagnetic radiation, ellipticity, position angle, shape, and colour \citep{Djo87,Dre87,Sco91,Van09a,Van09b,Kor09,Kor12}. These analyses were successful in acquiring information regarding the structure and development of some of these galaxies. However, these studies have done little analysis of the spirality of galaxies that were classified as elliptical.

Studying the morphology of large datasets of galaxies have attracted significant attention in the past decade \citep{Con03, Abr03, Bal08, Sha09, Ban10, Hue11}, and was driven by the increasing availability of automatically acquired datasets such as the data releases of the Sloan Digital Sky Survey \citep{Yor00}. However, attempts to automatically classify faint galaxy images along the Hubble sequence have been limited by the accuracy and capability of computer learning classification systems, and did not provide results that met the needs of practical research \citep{Tho08,Lin08}. This contention led to the {\it Galaxy Zoo} \citep{Lin08} project, which successfully used a web-based system to allow amateur astronomers to manually classify galaxies acquired by SDSS \citep{Lin10}, and was followed by other citizen science ventures based on the same platform such as {\it Galaxy Zoo 2} \citep{Mas11}, {\it Moon Zoo} \citep{Joy11}, and {\it Galaxy Zoo Mergers} \citep{Wal10}.

While it has been shown that amateurs can classify galaxies to their basic morphological types with accuracy comparable to professional astronomers \citep{Lin08}, manual classification may still be limited to what the human eye can sense and the human brain can perceive. For instance, the human eye can sense only 15 to 25 different levels of gray, while machines can identify 256 gray levels in a simple image with eight bits of dynamic range. The inability of the human eye to differentiate between gray levels can make it difficult to sense spirality in cases where the arms are just slightly brighter than their background, but not bright enough to allow detection by casual inspection of the galaxy image. In fact, this limitation might affect professional astronomers as much as it affects citizen scientists.

Since the human eye can only sense the crude morphology of galaxies along the Hubble sequence, and since the classification of galaxies is normally done manually, morphological classification schemes of galaxies are based on few basic morphological types. However, as these schemes are merely an abstraction of galaxy morphology, some galaxies can be difficult to associate with one specific shape, and many in-between cases can exist.


Here we use the Ganalyzer method to transform the galaxy images into their radial intensity plots \citep{Sha11}, and analyze the spirality of galaxies classified manually as elliptical and S0 by the {\it Galaxy Zoo}, RC3, and NA10 catalogues.

\section{Image analysis method}
\label{methodology}

The method that was used to measure the spirality of the galaxies in the dataset is the Ganalyzer method \citep{Sha11, Sha11b}. Unlike other methods that aim at classifying a galaxy into one of several classes of broad morphological types \citep{Abr03,Con03,Bal08,Sha09,Ban10,Hue11}, Ganalyzer measures the slopes of the arms to determine the spirality of a galaxy. Ganalyzer is a model-driven method that analyzes galaxy images by first separating the object pixels from the background pixels using the Otsu graylevel threshold \citep{Ots79}. The centre coordinates of the object are determined by the largest median value of the 5$\times$5 shifted window with a distance less than $0.1/\sqrt{S \over \pi}$ from the mass centre, where S is the surface area \citep{Sha11,Sha12}. This method allows the program to determine the maximum radial distance from the centre to the outermost point, as well as the major and minor axes by finding the longest distance between two points which pass through the centre for the major axis, and then assigning the perpendicular line as the minor axis \citep{Sha11}. The ellipticity is defined as the ratio of the lengths of the minor axis to the major axis \citep{Sha11}. Comparison of the ellipticity of 1000 galaxies to the ellipticity computed by SDSS  (using isoA and isoB) shows a high Pearson correlation of $\sim$0.93 between the two measurements.

\begin{figure}[ht]
\centering
\includegraphics[scale=0.5]{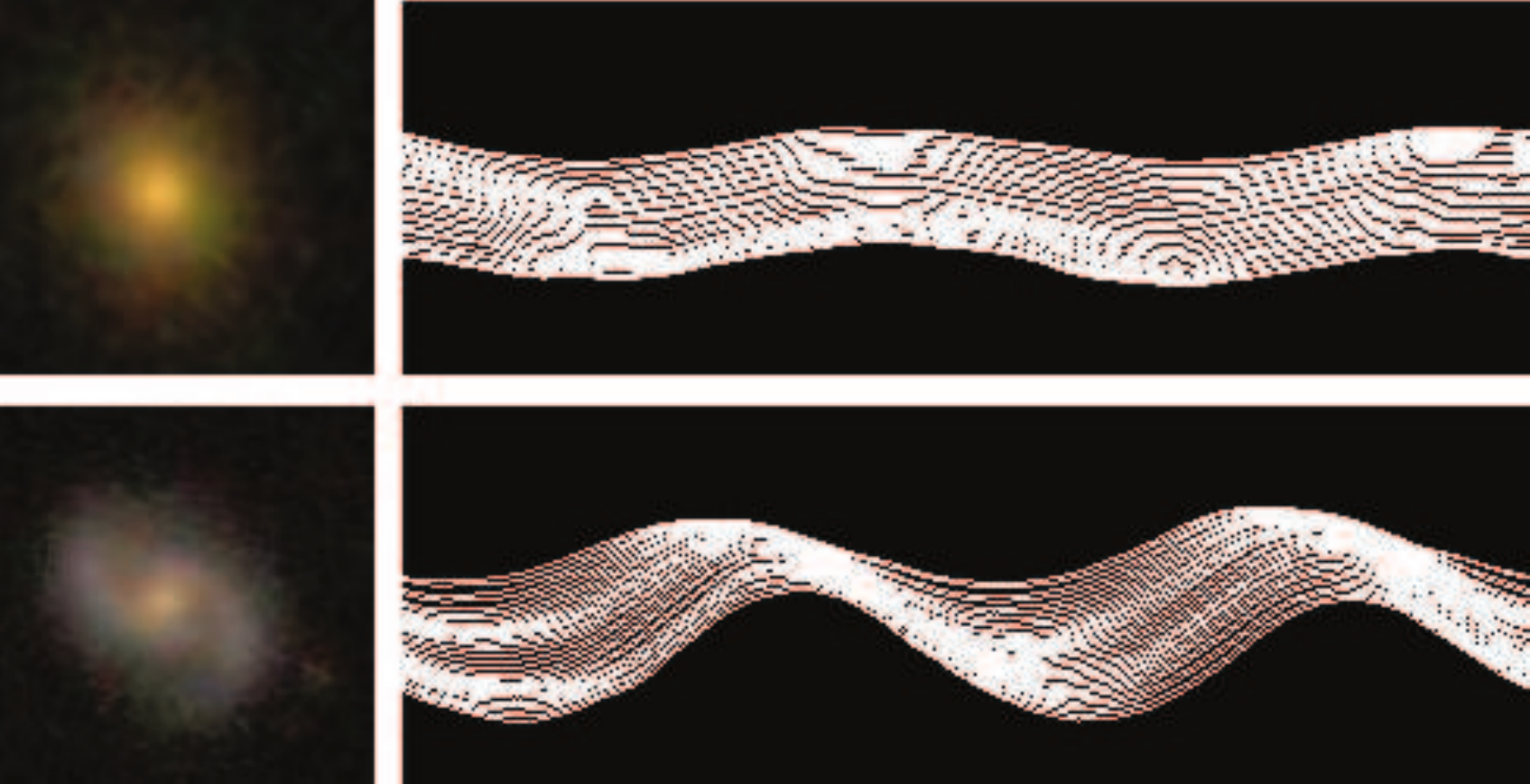}
\caption{Galaxy images and their transformation to radial intensity plots such that the Y axis is the pixel intensity and the X axis is the polar angle}
\label{radial}
\end{figure}

After the centre coordinates of the galaxy $O_x,O_y$ and the radius $r$ are determined, the galaxy is transformed into its radial intensity plot such that the intensity value of the pixel $(x,y)$ in the radial intensity plot is the intensity of the pixel at coordinates $(O_x + r\sin{\theta}, O_y-r\cos{\theta})$ in the original galaxy image, such that $\theta$ is a polar angel of [0,360], and $r$ is the radial distance that ranges from 0.4 to 0.75 of the galaxy radius, producing an image of dimensionality of 360$\times$35 \citep{Sha11,Sha12}. Figure~\ref{radial} shows an example of two galaxies and their transformation such that the Y axis is the pixel intensity and the X axis is the polar angle. 

As the figure shows, in the case of the elliptical galaxy the peaks are aligned on the same vertical line, while in the case of the spiral galaxy the peaks shift. The spirality is then measured by the slope of the groups peaks as described in \citep{Sha11}, such that the peak in radial distance {\it r} is grouped with the peak in radial distance {\it r}+1 if the difference between their polar angles is less than 5$^o$. This transformation makes it easier for machines to detect and measure the spirality, but can also detect spirality in galaxies that might look to the human observer as elliptical since the human eye can only recognize 15-25 gray levels, making it difficult to notice subtle spirality when looking at a raw galaxy image. For instance, Tables 1 and 2 
show several SDSS galaxy images classified manually by {\it Galaxy Zoo} participants as elliptical, with their radial intensity plot transformation and their spirality as measured by Ganalyzer. To test how the method analyzes tidally disrupted elliptical galaxies \citep{Dok05}, we used several tidally disrupted galaxies  from the NA10 catalogue, displayed in Table 3.

\begin{table*}[ht]
 \centering
 \begin{minipage}{140mm}
  \caption{Sample SDSS galaxy images with the Otsu binary transform, radial intensity plot transforms and the measured spirality}
  \label{galaxy_table}
  \begin{tabular}{@{}l|c|c|c@{}}
  \hline
     Galaxy image & Otsu transform & Radial Intensity Plot & Spirality  \\
  \hline
  
\includegraphics[scale=0.75]{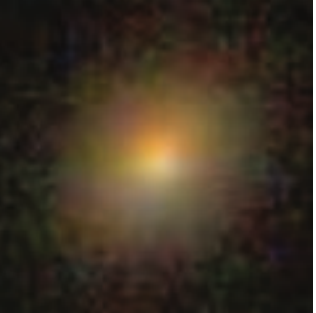} & \includegraphics[scale=0.52]{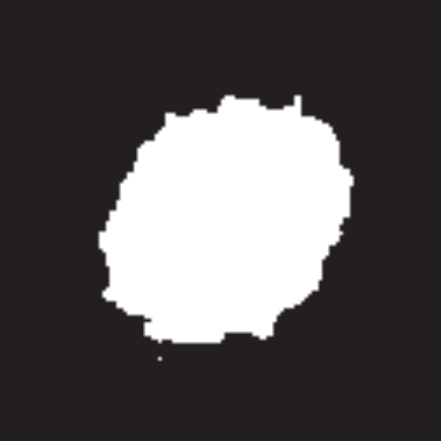} & \includegraphics[scale=0.75]{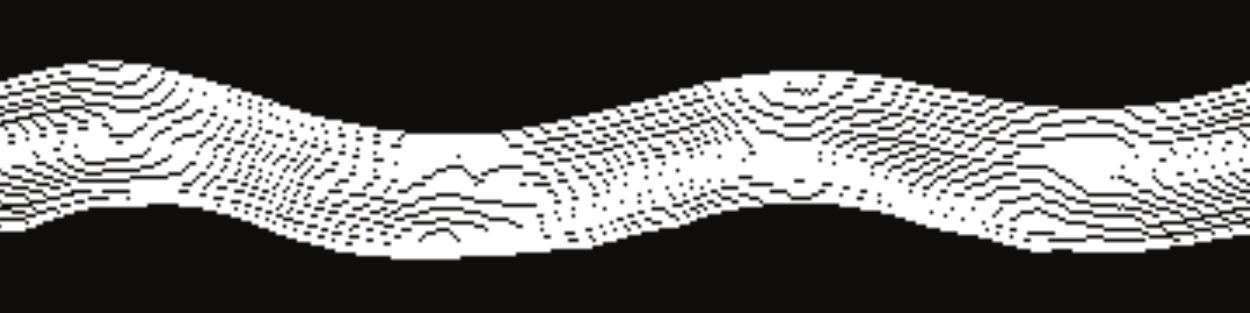} & 0 \\

\includegraphics[scale=0.75]{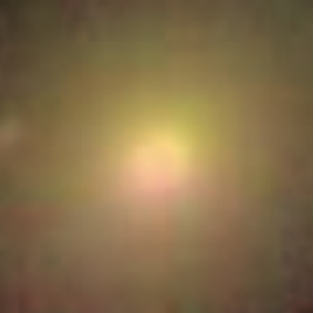} & \includegraphics[scale=0.68]{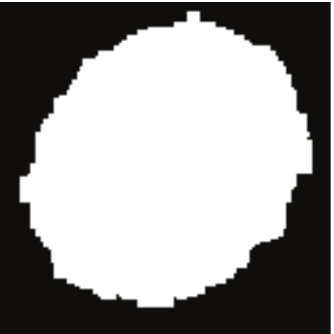} & \includegraphics[scale=0.75]{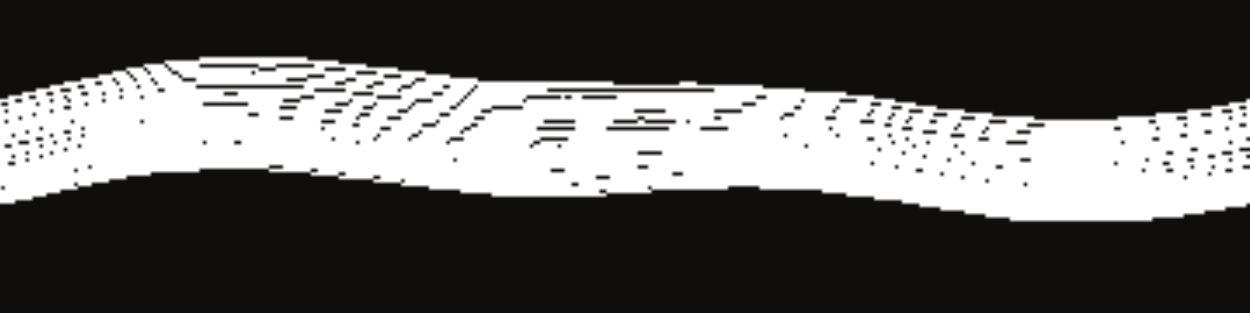} & 0 \\

\includegraphics[scale=0.75]{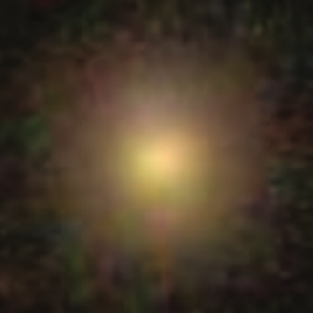} & \includegraphics[scale=0.68]{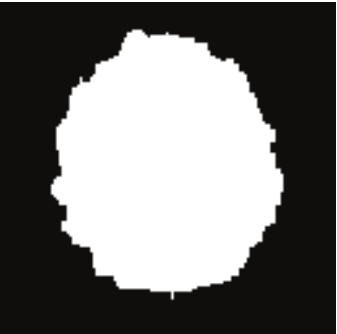} & \includegraphics[scale=0.75]{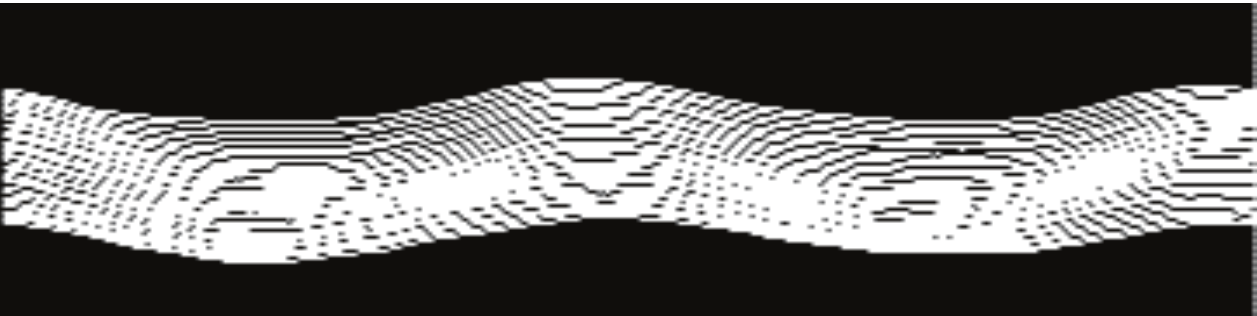} & 0 \\

\includegraphics[scale=0.75]{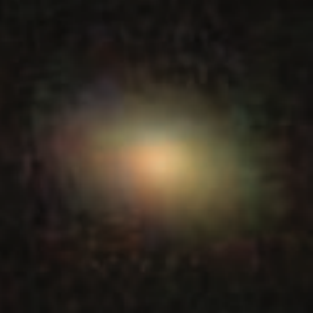} & \includegraphics[scale=0.52]{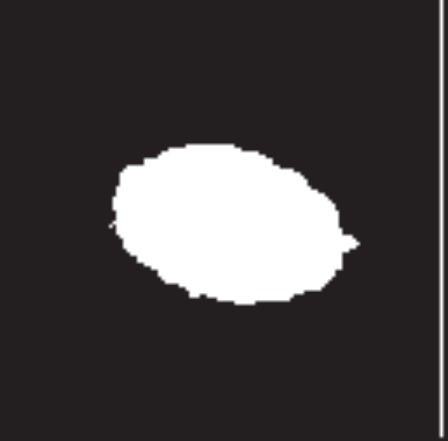} & \includegraphics[scale=0.75]{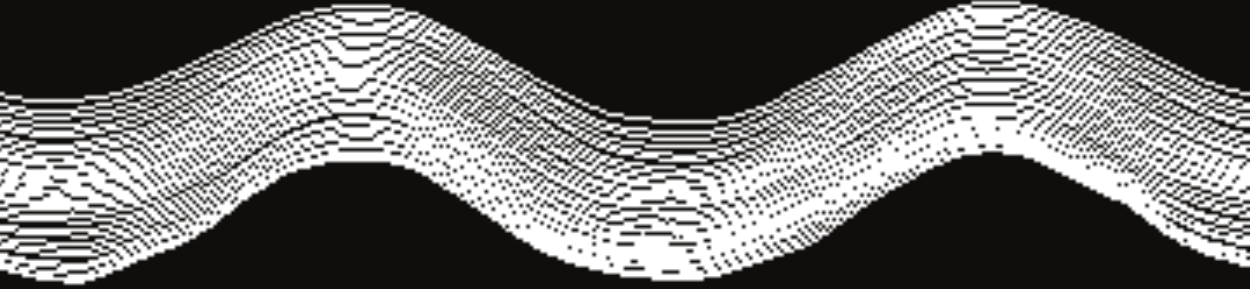} & 0.06 \\

\includegraphics[scale=0.75]{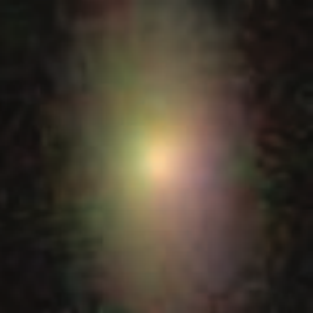} & \includegraphics[scale=0.52]{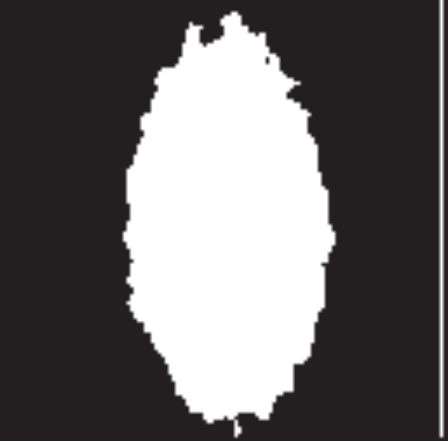} & \includegraphics[scale=0.75]{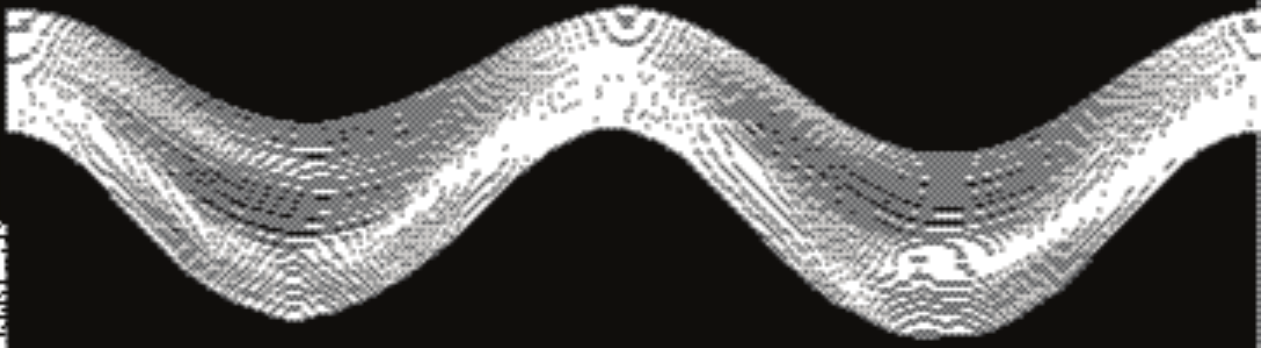} & 0.14 \\

\includegraphics[scale=0.75]{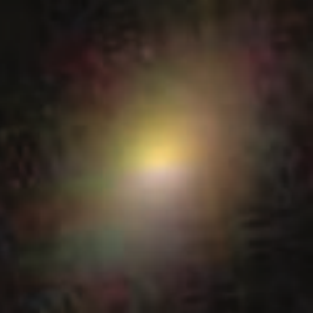} & \includegraphics[scale=0.52]{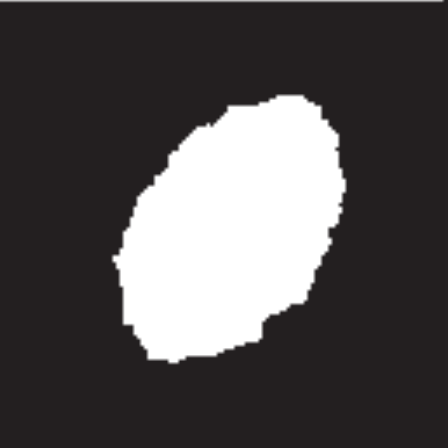} & \includegraphics[scale=0.75]{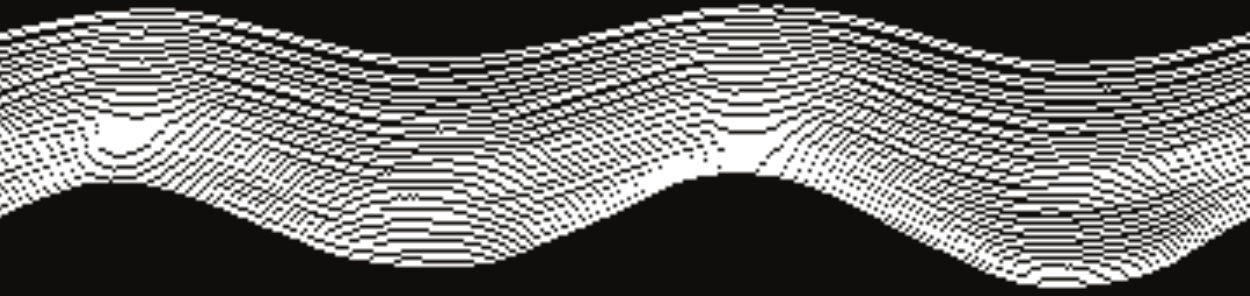} & 0.21 \\

\hline
\end{tabular}
\end{minipage}
\end{table*}

\begin{table*}[ht]
 \centering
 \begin{minipage}{140mm}
  \caption{Sample SDSS galaxy images with the Otsu binary transform, radial intensity plot transforms, and the measured spirality}
  \label{galaxy_table2}
  \begin{tabular}{@{}l|c|c|c@{}}
  \hline
     Galaxy image & Otsu transform & Radial Intensity Plot & Spirality  \\
  \hline
 
\includegraphics[scale=0.75]{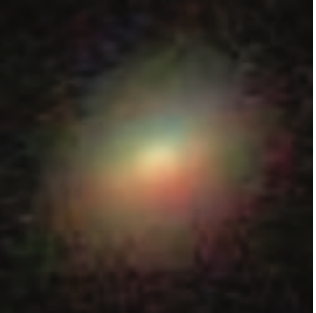} & \includegraphics[scale=0.52]{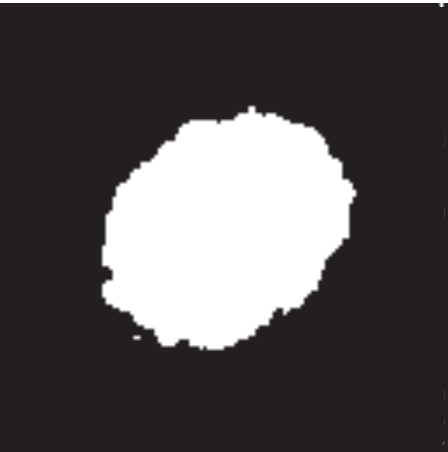} & \includegraphics[scale=0.75]{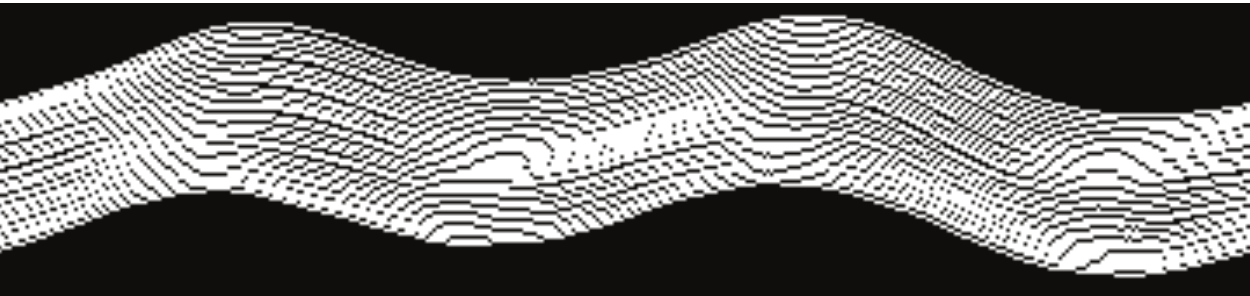} & 0.44 \\

\includegraphics[scale=0.75]{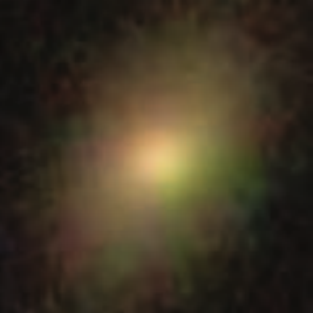} & \includegraphics[scale=0.52]{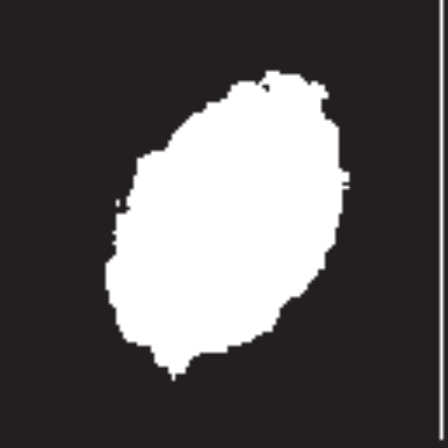} & \includegraphics[scale=0.75]{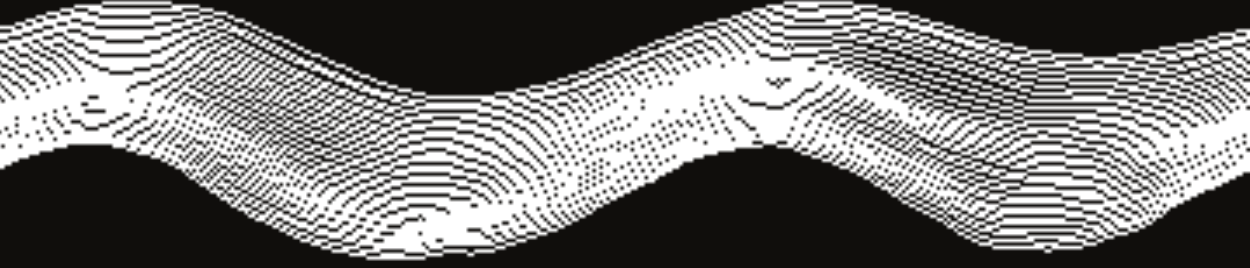} & 0.53 \\

\includegraphics[scale=0.75]{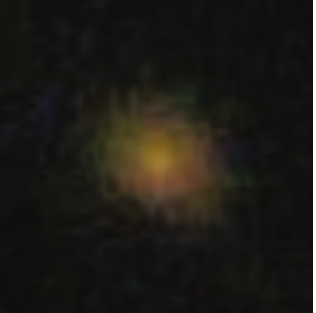} & \includegraphics[scale=0.52]{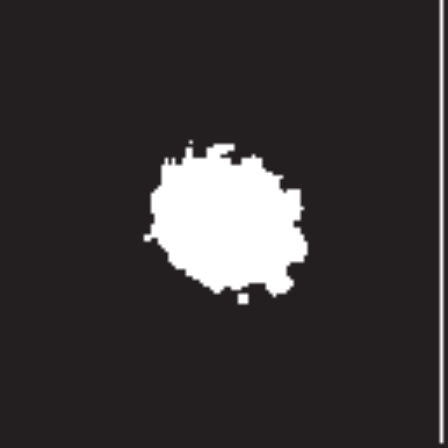} & \includegraphics[scale=0.75]{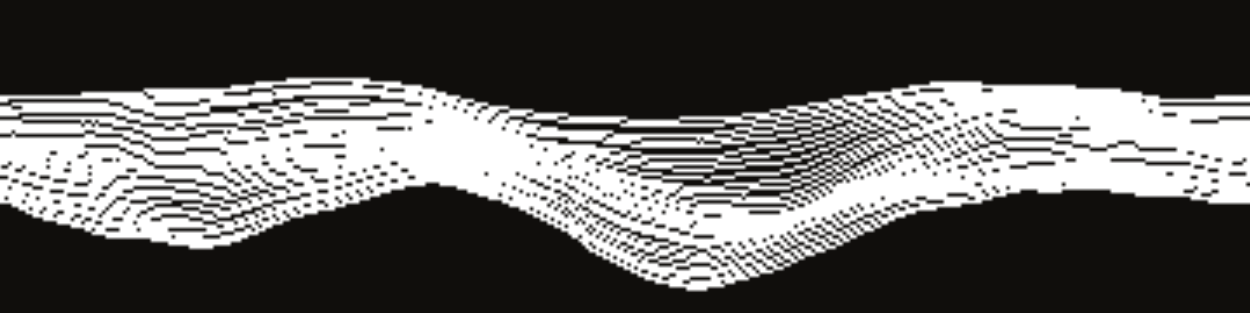} & 0.60 \\

\includegraphics[scale=0.75]{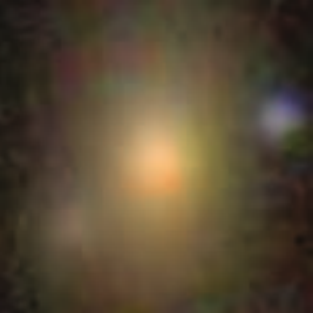} & \includegraphics[scale=0.52]{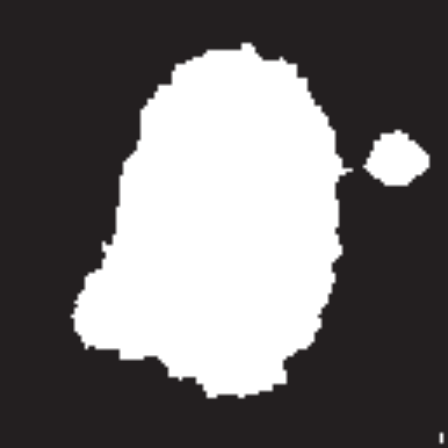} & \includegraphics[scale=0.75]{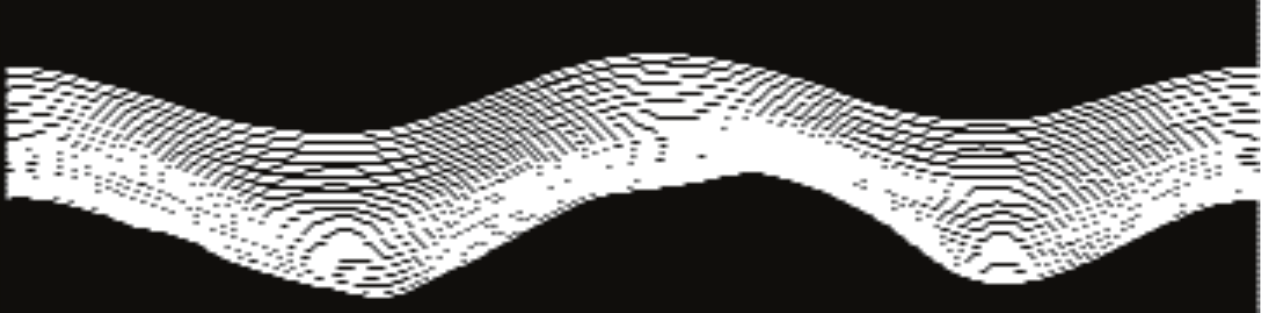} & 0.77 \\

\includegraphics[scale=0.75]{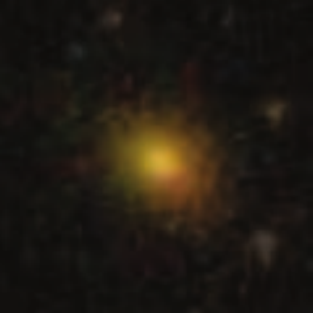} & \includegraphics[scale=0.52]{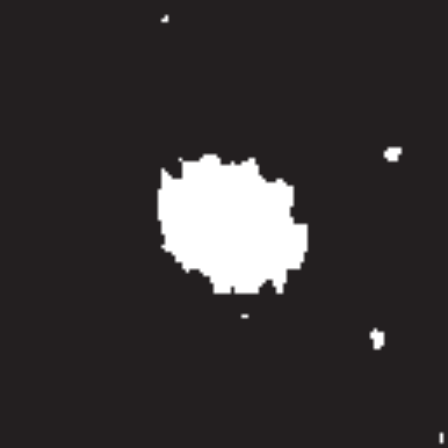} & \includegraphics[scale=0.75]{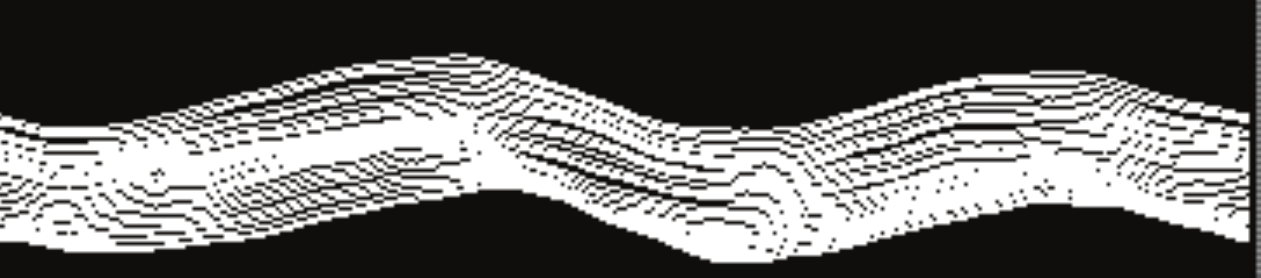} & 1.54 \\

\hline
\end{tabular}
\end{minipage}
\end{table*}

\begin{table*}[ht]
 \centering
 \begin{minipage}{140mm}
  \caption{Sample of tidally disrupted galaxy images taken from NA10 catalogue}
  \label{galaxy_table3}
  \begin{tabular}{@{}l|c|c|c@{}}
  \hline
     Galaxy image & Otsu transform & Radial Intensity Plot & Spirality  \\
  \hline

\includegraphics[scale=0.35]{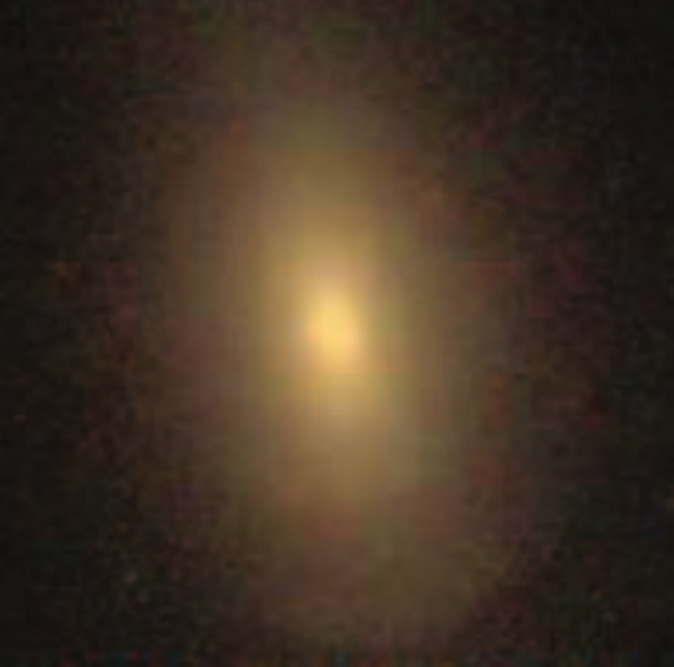} & \includegraphics[scale=0.35]{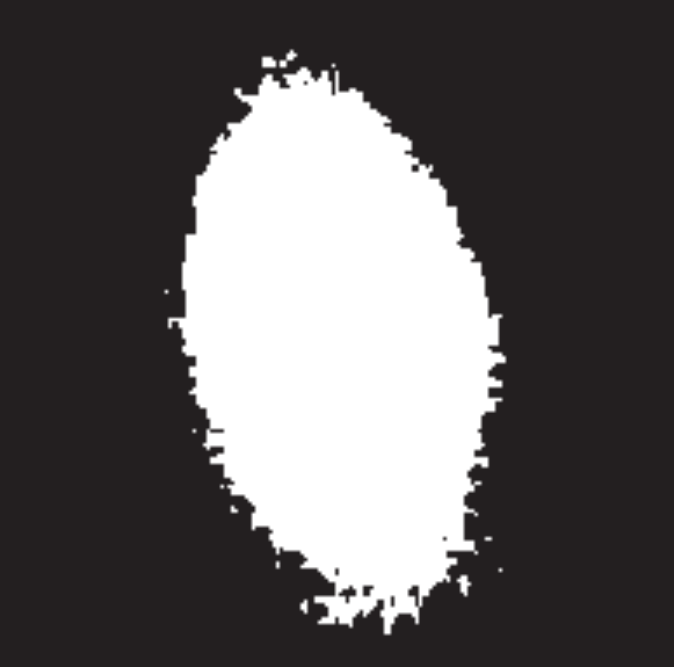} & \includegraphics[scale=0.75]{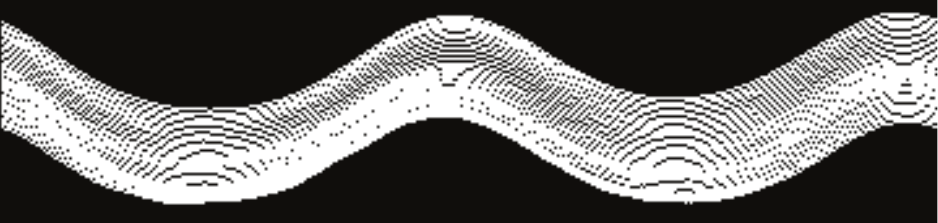} & 0.12 \\
 
\includegraphics[scale=0.30]{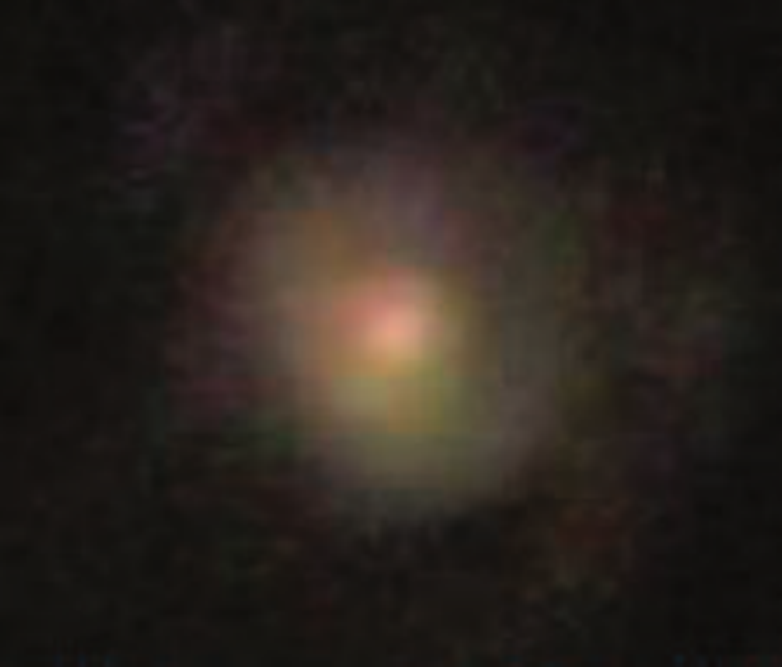} & \includegraphics[scale=0.30]{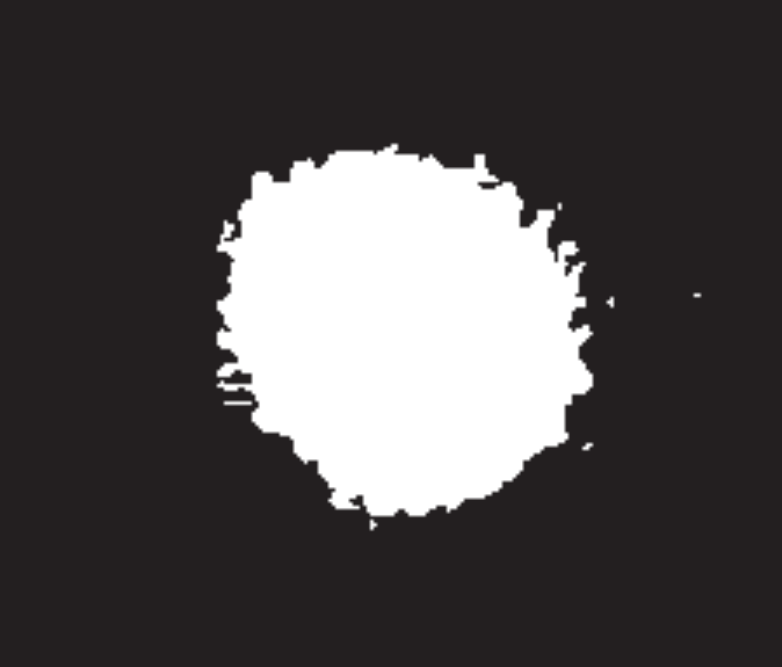} & \includegraphics[scale=0.75]{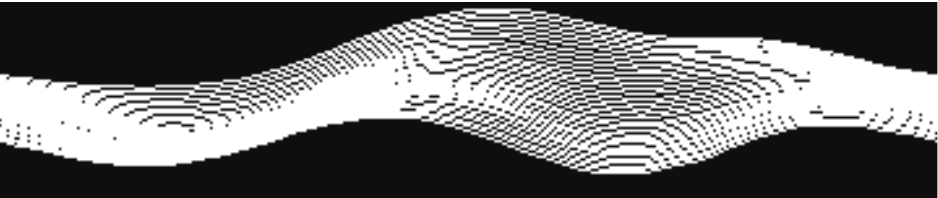} & 1.05 \\

\includegraphics[scale=0.32]{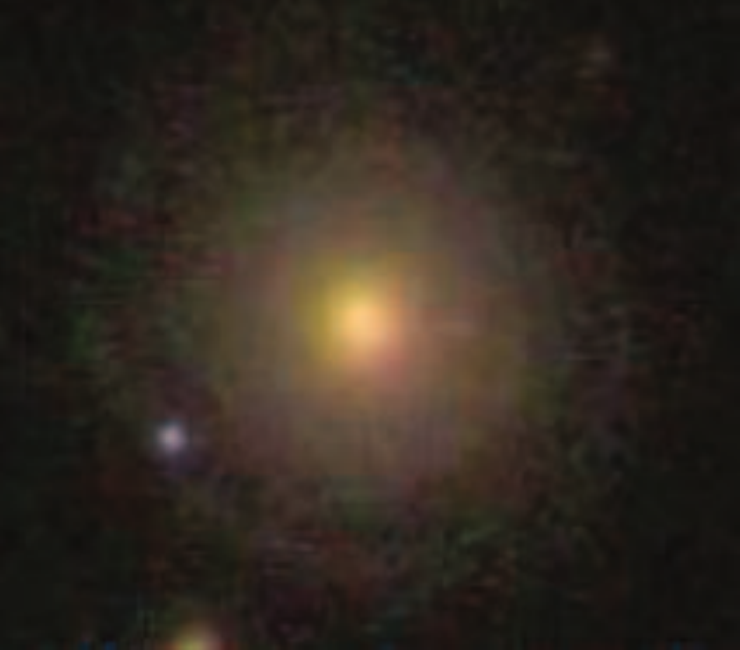} & \includegraphics[scale=0.30]{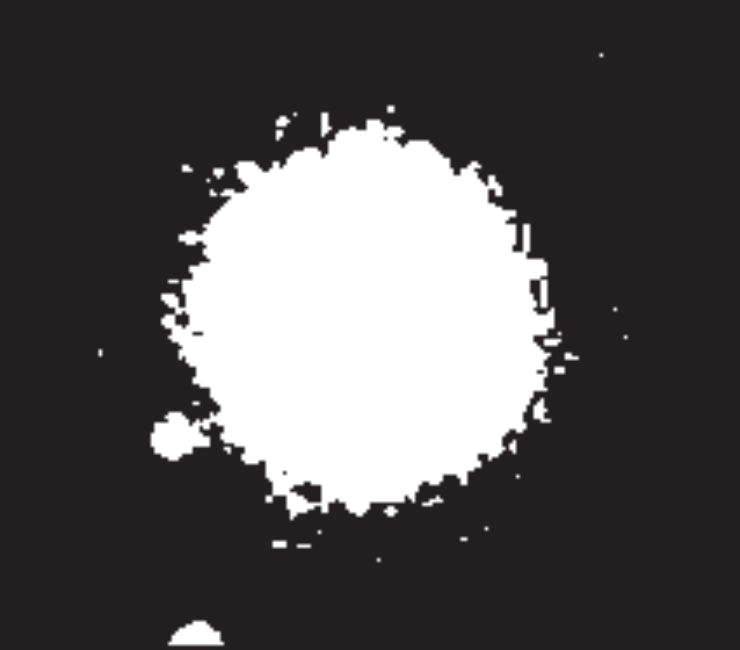} & \includegraphics[scale=0.75]{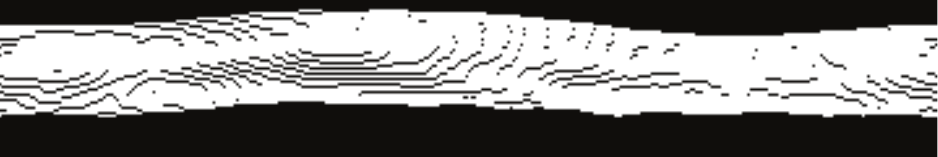} & 0 \\

\includegraphics[scale=0.35]{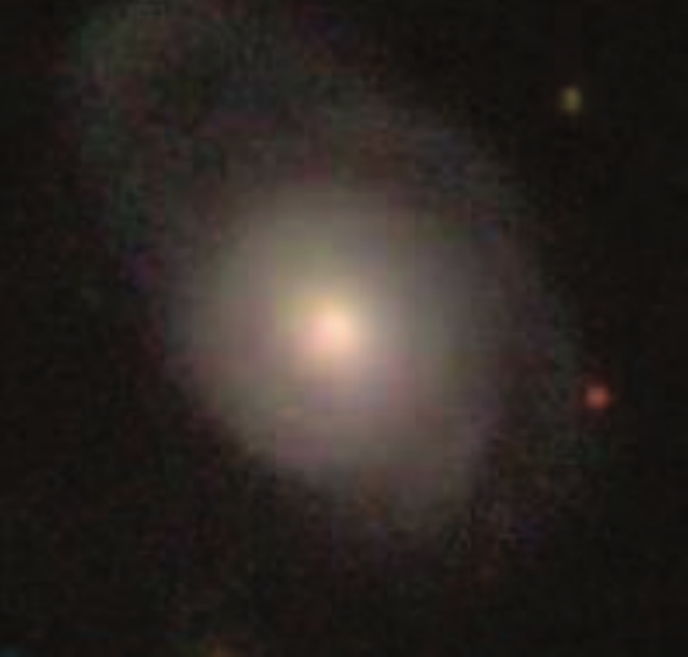} & \includegraphics[scale=0.35]{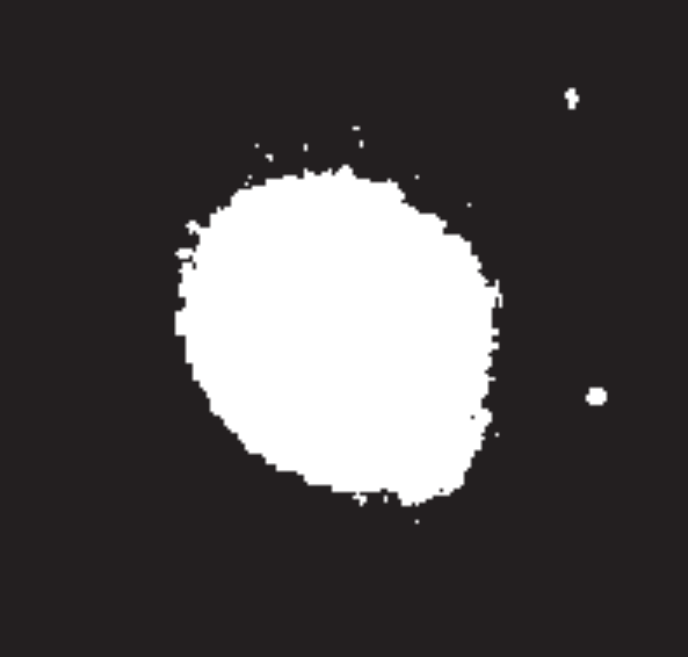} & \includegraphics[scale=0.75]{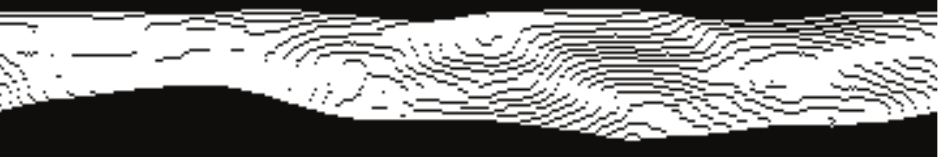} & 0 \\

\includegraphics[scale=0.32]{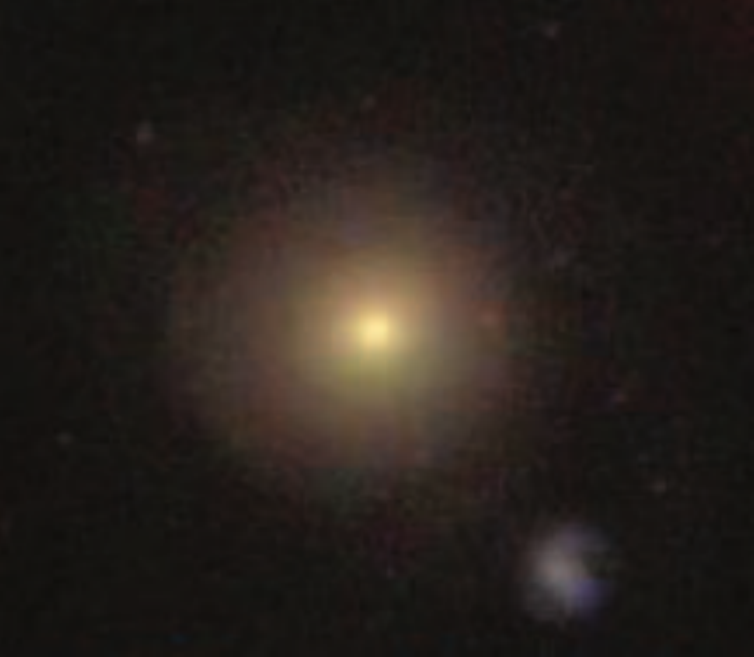} & \includegraphics[scale=0.32]{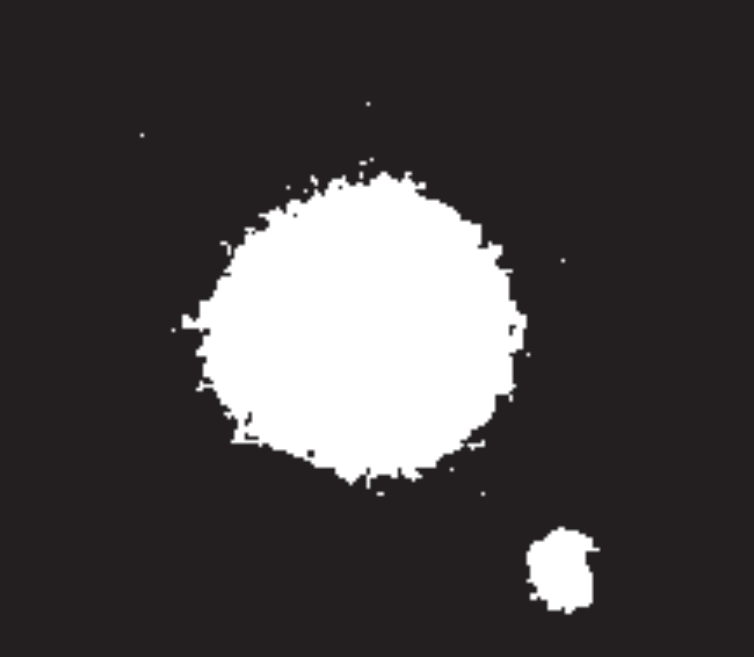} & \includegraphics[scale=0.75]{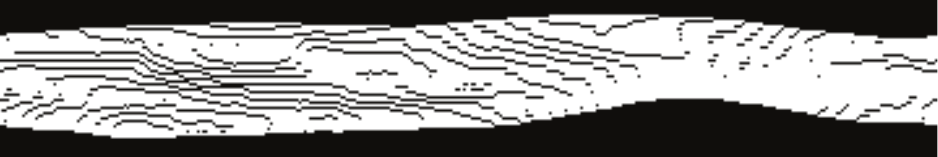} & 0 \\

\includegraphics[scale=0.32]{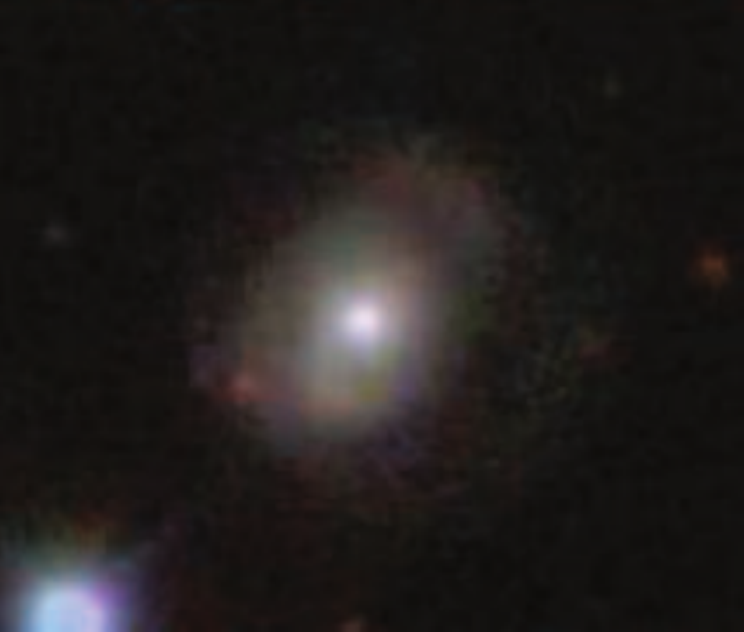} & \includegraphics[scale=0.32]{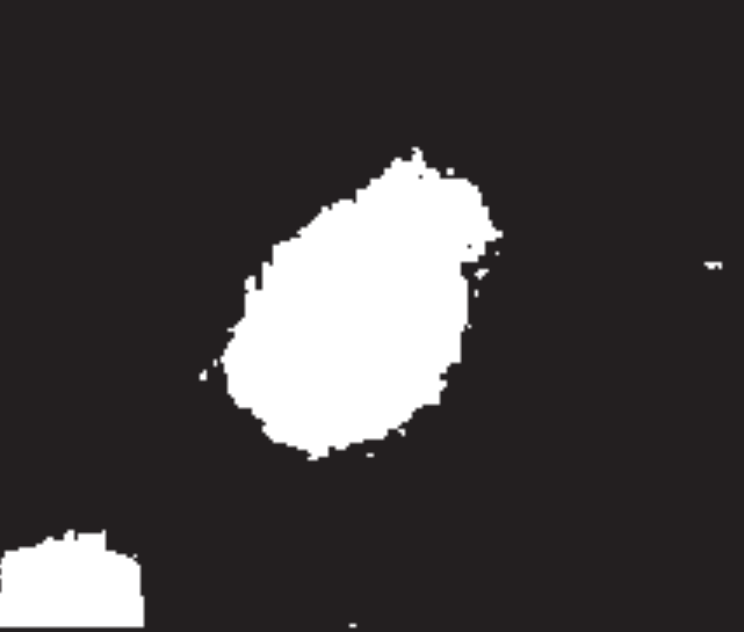} & \includegraphics[scale=0.75]{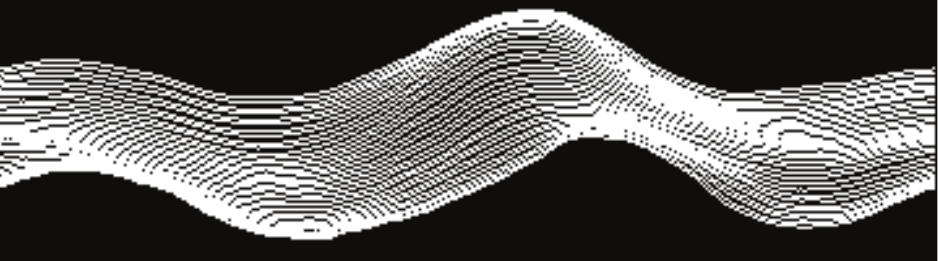} & 1.14 \\

\hline
\end{tabular}
\end{minipage}
\end{table*}

If the radial intensity plot does not feature peaks the galaxy is defined as pure elliptical. Elliptical and lenticular galaxies in some cases can also have peaks in their radial intensity plot due to the position angle, but in these cases all peaks will be aligned on the same vertical line so that the slope will be very close to zero, and therefore the galaxy will be identified as elliptical. An exception can be in cases of S0 galaxies in which the position angle of the disk is different from the position angle of the galaxy, but the difference is not greater than 5$^o$. In that case Ganalyzer might consider the disk and the galaxy as the same arm, but the difference in the position angles will lead to a certain slope in that arm. Therefore, the arms of the galaxy will have a certain slope when measured using Ganalyzer.

The radial intensity plot can allow the detection of subtle curves in the arms that might not be easily detected by manual observation of the raw galaxy image, but becomes noticeable in its radial intensity plot. Therefore, it is possible that many galaxies that were classified manually as elliptical might in fact feature a certain spirality \citep{Sha11}. As the table shows, while the galaxies seem elliptical to the unaided human eye, the radial intensity plot transformations of the galaxies show that the peaks of maximal intensity shift, meaning that these galaxies feature certain curves in the arms.

By defining spirality and ellipticity thresholds Ganalyzer can also be used for classifying galaxies into their broad morphological types of elliptical, spiral and edge-on, and a thorough discussion and experimental results about galaxy classification with Ganalyzer are described in \citep{Sha11}. In previous experiments with Ganalyzer \citep{Sha11,Sha11b,Sha12} thresholds were applied to the slopes in the radial intensity plots so that the decision whether a galaxy is spiral or not is in agreement with the perception of a person observing the raw galaxy image. However, as described above, in this study Ganalyzer is not used as a classifier, but as a tool to measure and detect the existence of galaxy spirality. Since the radial intensity plot provides a more sensitive view of galaxy spirality than the non-transformed raw image, no thresholds are used in this study in order to utilize the ability of the radial intensity plots to detect subtle slopes in the galaxy arms and to test whether galaxies that seem elliptical to the human eye are indeed ellipticals, or have a subtle spirality that is difficult to measure using the unaided eye. That is, the purpose of the method described in this section is not to mimic the human eye, but test whether the human eye observing the raw galaxy image is indeed the most accurate tool to determine whether a galaxy is spiral or elliptical.

\section{Data}
\label{data}

The data used in the experiment are galaxies acquired by Sloan Digital Sky Survey, and were classified manually by the participants of the Galaxy Zoo project \citep{Lin08,Lin10}. All galaxies in the dataset have redshift, and the classification results were based on the corrected {\it super clean} dataset described in \citep{Lin10}. For the study, only galaxies that were classified by Galaxy Zoo participants as ellipticals were used, and the dataset consisted of 60,518 galaxies. The images were downloaded automatically by using the CAS server. The galaxies were also divided into six bins based on their redshift, ranges from 0 to 0.3, such that each bin had a redshift range of 0.05. The number of galaxies in each redshift is specified in Table 4.

\begin{table}
 \centering
 \begin{minipage}{70mm}
  \caption{Percent of galaxies with spirality greater than zero, based on redshift}
  \label{redshift_dist}
  \begin{tabular}{@{}llc|c@{}}
  \hline
  Redshift & \# of galaxies \\
  \hline
  0.00 to 0.05 &  7767           \\
  0.05 to 0.10 &   12248      \\
  0.10 to 0.15 &   12107       \\
  0.15 to 0.20 &   14451        \\
  0.20 to 0.25 &   10088       \\
  0.25 to 0.30 &   2858         \\
 \hline
\end{tabular}
\end{minipage}
\end{table}

The efficiency of Ganalyzer can be affected by two or more galaxies that appear very close to each other in the image, either due to merging or superpositioning. Since one galaxy can be segmented with part or all of the other galaxy, Ganalyzer might detect the other galaxy as an arm. In most cases such ``arm" is not expected to be mistakenly identified as sharp spirality because the angle of the brightest point compared to the center is not expected to shift, but it can lead to the false detection of mild spirality that is not based on the morphology of the target galaxy. To avoid analyzing overlapping galaxies each image was scanned for PSFs as done in \citep{Sha05a,Sha05b}, and when more than one PSF is detected the image is ignored. Out of the galaxies classified by Galaxy Zoo as elliptical $\sim$12.05\% were detected as galaxies with more than one nucleus and were therefore rejected from the analysis.

Other catalog that were used in this study were the RC3 catalog \citep{Cor94}, of which 261 galaxies classified as ellipticals and 640 galaxies classified as S0 were used, and the NA10 catalog \citep{Nair2010}, of which 2705 galaxies that were classified as ellipticals, and 1964 galaxies that were classified as S0 were used in the experiment. Additionally, 7638 galaxies of the NA10 catalog classified as spirals were also used in the analysis.

\section{Results}
\label{results}

\begin{figure}
\centering
\includegraphics[scale=0.35]{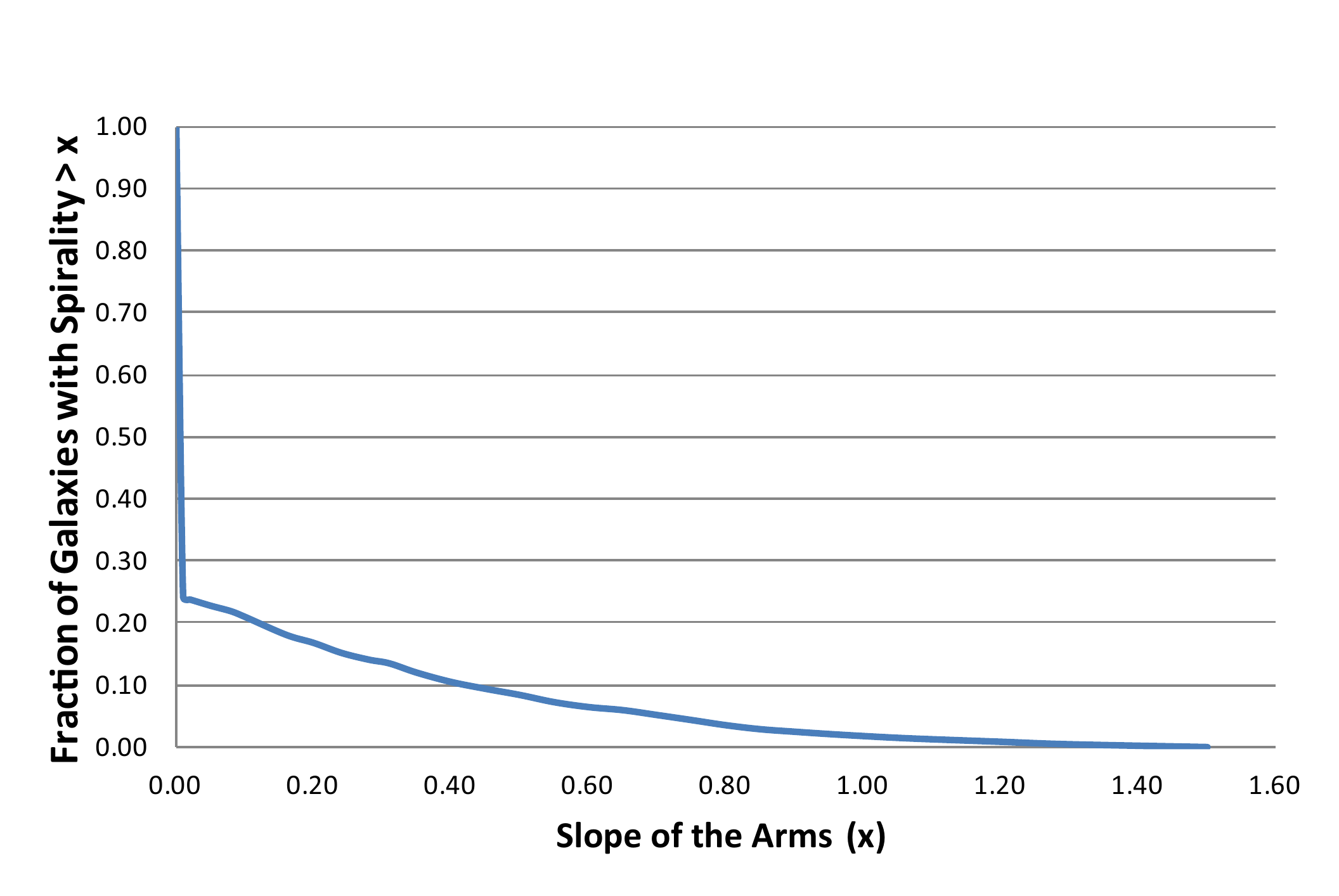}
\caption{Slope of the arms of the entire galaxy population of $\sim$60,000 galaxies classified as ellipticals}
\label{spirality_distribution}
\end{figure}

Figure~\ref{spirality_distribution} shows the distribution of the slopes of the arms of the galaxies classified manually as elliptical. As the figure shows, $\sim$24\% of the galaxies exhibit nonzero signal for spirality, and $\sim$10\% of the galaxies had a slope of the arms greater than 0.4, indicating that many of the {\it Galaxy Zoo} galaxies that were classified manually as ellipticals actually have some spirality. Expectedly, the fraction of galaxies that meet the spirality threshold decreases as the slope of the arms gets larger, and just less than 2\% of the galaxies that were classified manually as ellipticals were detected to have a measured slope greater than 1. As the graph shows, the measured slope of the arms of most galaxies is close to 0.

\begin{figure}
\centering
\includegraphics[scale=0.60]{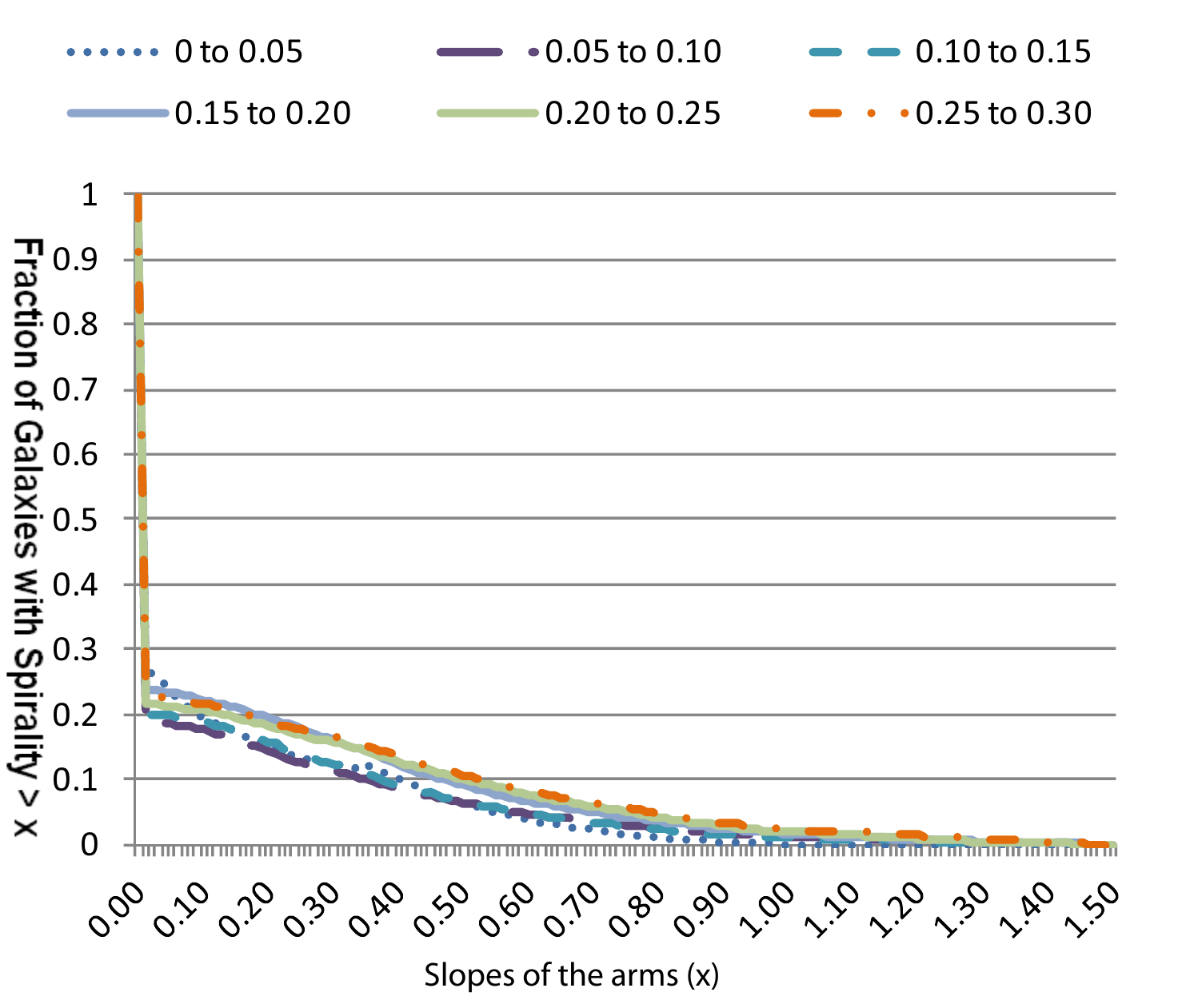}
\caption{The distribution of the slopes of the arms in galaxies of different redshifts ranges classified by Galaxy Zoo as ellipticals}
\label{redshift_0_015}
\end{figure}

Figure~\ref{redshift_0_015} shows the distribution of the slopes of the arms in a galaxy population of different redshifts. As can be learned from the Figure, the slope of the arms increases with the redshift, showing that at higher redshifts human readers find it more difficult to detect spirality by eye and tend to classify more galaxies as ellipticals. Correlating the apparent magnitude of these galaxies with the slope of the arms provided a weak Pearson correlation value of -0.036, showing that the analysis is merely weakly dependent on apparent magnitude in the redshift range used in this study.

\begin{figure}
\centering
\includegraphics[scale=0.28]{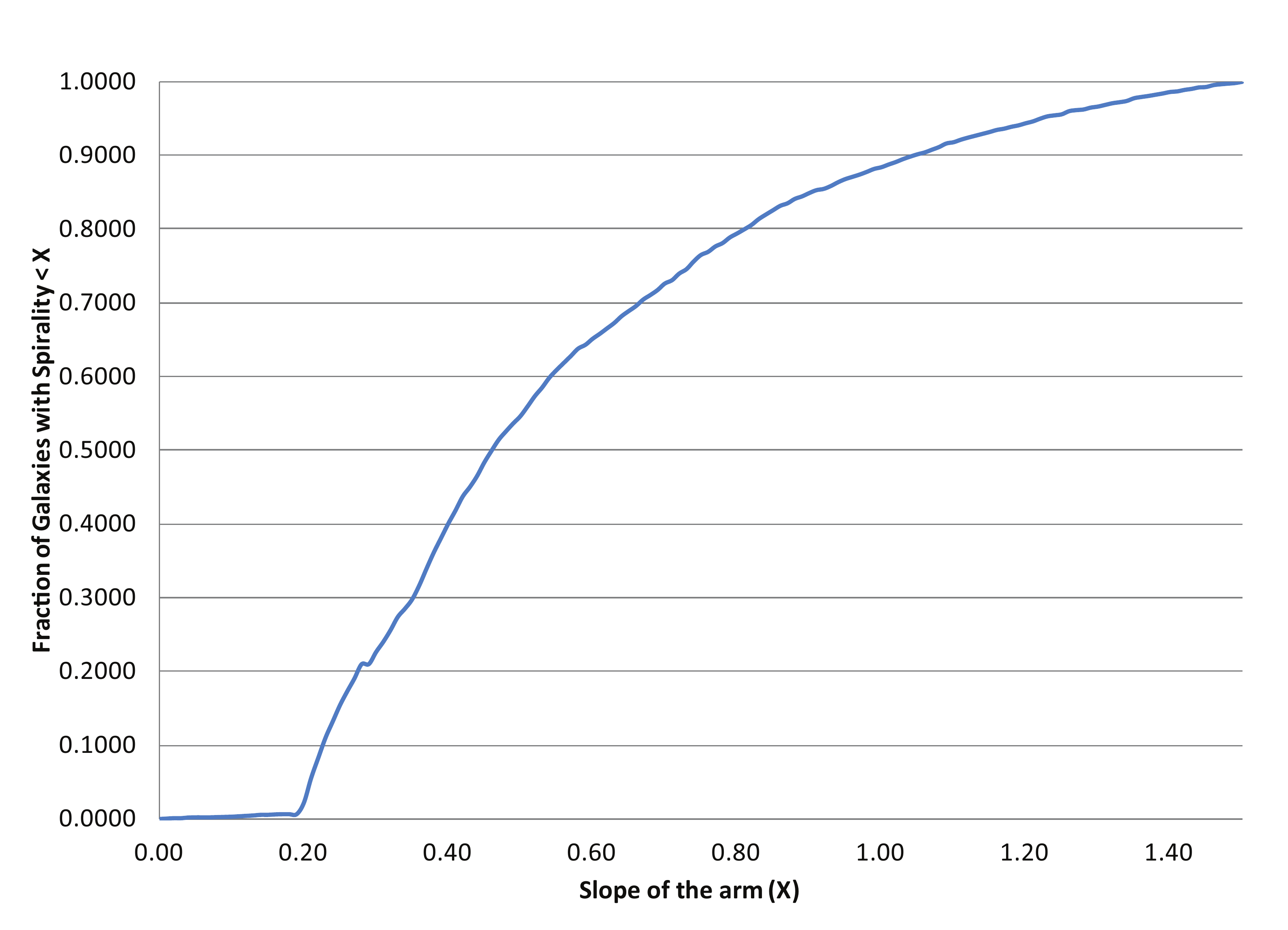}
\caption{The distribution of the slopes of the arms in galaxies classified by Galaxy Zoo participants as spiral galaxies}
\label{gz_spirals}
\end{figure}

While the experiments above show that human readers can in some cases fail to notice mild spirality, we also tested the error rate of human readers when they determine that the galaxy that they observe is spiral. Figure~\ref{gz_spirals} shows the distribution of the measured slope of the arms among galaxies that were classified by Galaxy Zoo participants as spiral. As the figure shows, when the human eye is able to detect spirality, spirality does exist, and galaxies classified by human observers as spirals are rarely galaxies that do not have any spirality in them. The reason could be the limited sensitivity of the human eye in detecting spirality, so that once spirality can be detected by the human eye it is above a certain spirality threshold that can be sensed by applying the analysis of the radial intensity plot of the galaxy as described in Section~\ref{methodology}.

\begin{figure}
\centering
\includegraphics[scale=0.39]{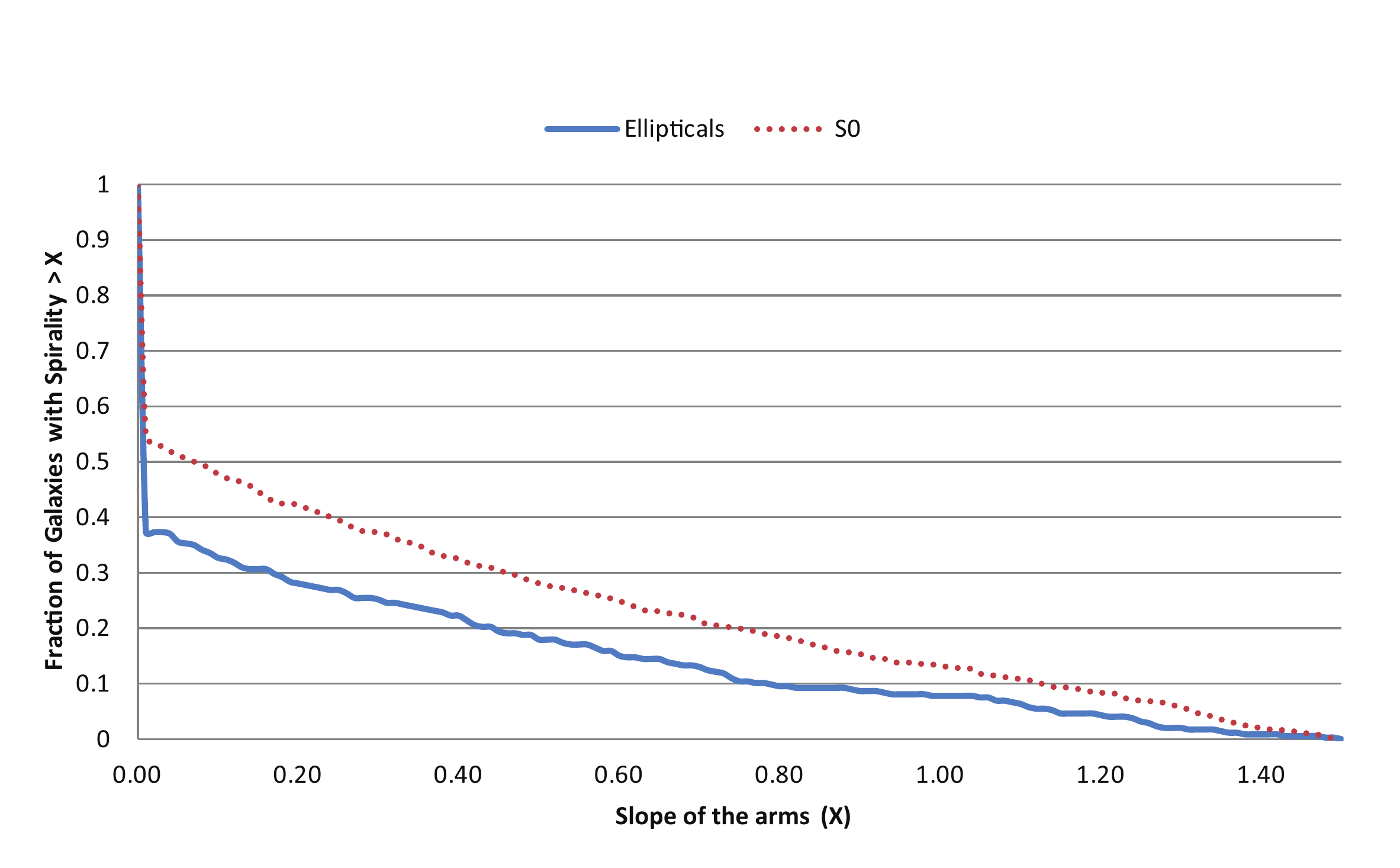}
\caption{The distribution of the slopes of the arms in galaxies that were classified by RC3 as ellipticals and S0 galaxies}
\label{rc3}
\end{figure}

The results obtained with the {\it Galaxy Zoo} data were compared also to the analysis using data taken by the RC3 catalogue \citep{Cor94} and the NA10 catalogue \citep{Nair2010}. Unlike {\it Galaxy Zoo} that was classified by amateurs, RC3 and NA10 were both classified by professional astronomers. Figure~\ref{rc3} shows the slopes of the arms of galaxies classified as ellipticals and S0 by RC3.

As the graph shows, galaxies classified as S0 have a higher arm slopes compared to galaxies classified as ellipticals. The figure also shows that like the {\it Galaxy Zoo} catalog, the arms of many of the galaxies classified by professional astronomers as ellipticals also has a certain slope. These results are in agreement with the observation of \cite{Lin08}, according which professional astronomers do not outperform amateur astronomers in classification of galaxies into their broad morphological types.

\begin{figure}
\centering
\includegraphics[scale=0.39]{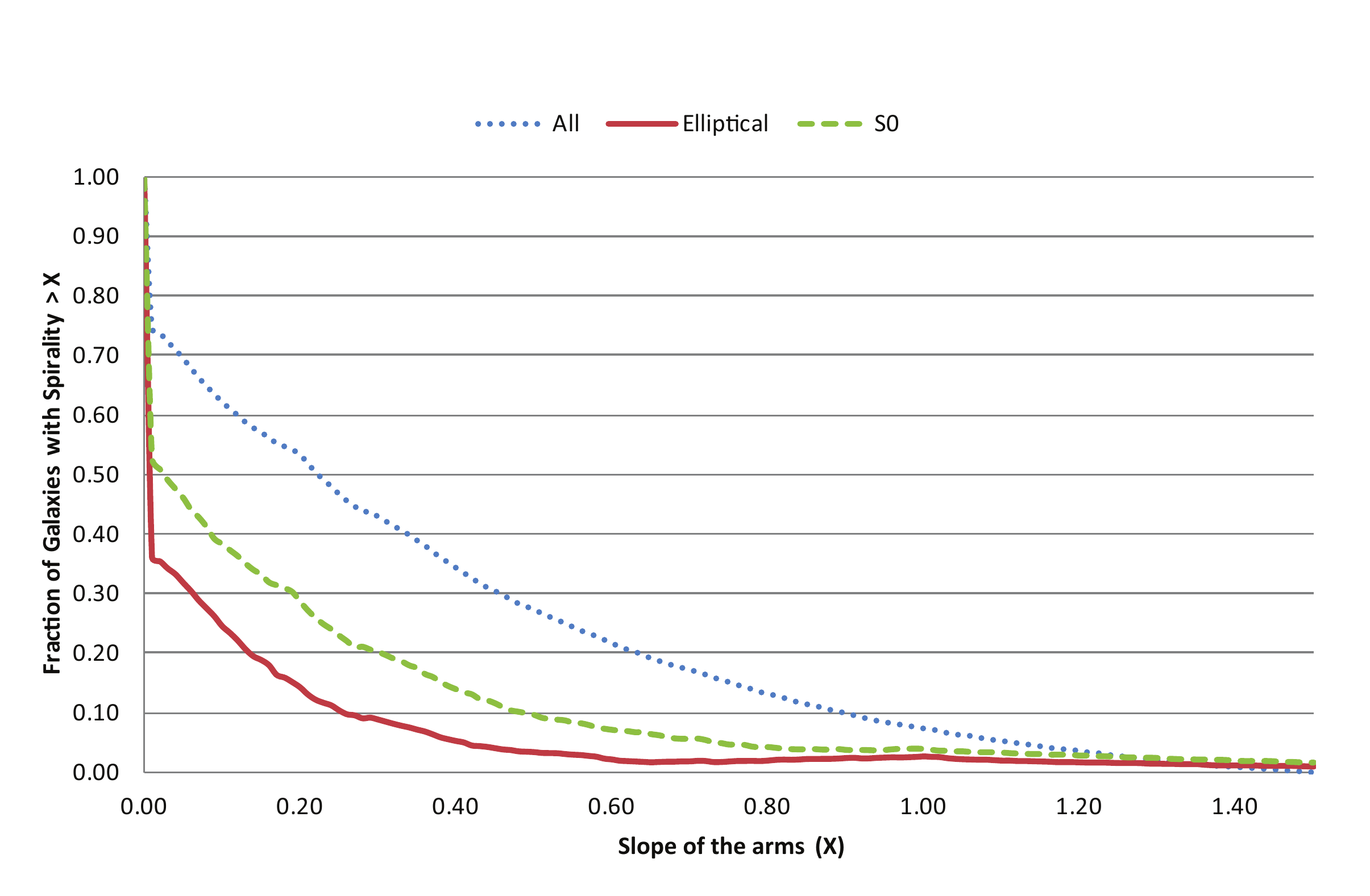}
\caption{The distribution of the slopes of the arms in galaxies that were classified by NA10 as ellipticals and S0 galaxies}
\label{na10}
\end{figure}

Figure~\ref{na10} shows the slopes of the arms of galaxies classified as ellipticals and S0 in the NA10 catalog. These results are in agreement with the results of the RC3 catalog, showing higher arm slopes in galaxies classified manually as S0.

\begin{figure}
\centering
\includegraphics[scale=0.38]{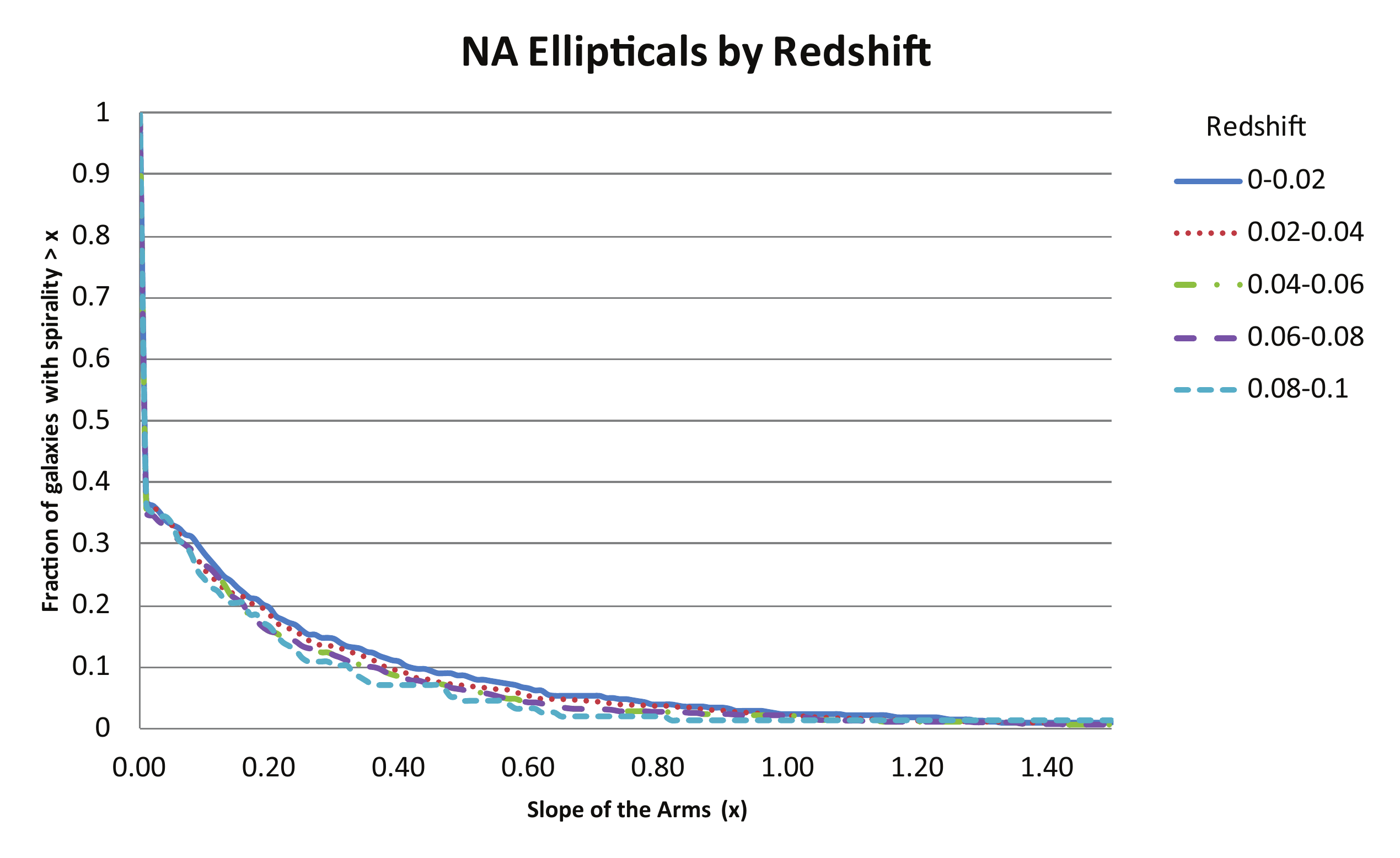}
\caption{The distribution of the slopes of the arms of galaxies that were classified by NA10 as ellipticals for different redshift ranges}
\label{na10_redshift}
\end{figure}

The distribution of the slopes of the arms was also analyzed for different redshifts, as displayed by Figure~\ref{na10_redshift},  showing galaxies classified as ellipticals. As the figure shows,the fraction of galaxies with non-zero slope classified as ellipticals in the NA10 catalog does not change significantly with the redshift in the tested redshift ranges. 

\section{Conclusions}
\label{conclusions}

In this study we used computer-aided analysis based on the radial intensity plots of SDSS galaxy images to examine spirality in galaxies that were classified manually as elliptical. While the unaided human eye provides a limited tool for analyzing elliptical galaxy images due to the limited sensitivity of the human vision to different gray levels, transforming the images to their radial intensity plots allows much easier detection of the spirality.

The results suggest that more than a third of the galaxies that were classified manually by {\it Galaxy Zoo} participants as elliptical actually have a certain spirality. Although in most cases the spirality was low, 10\% of the galaxies classified as elliptical had a slope greater than 0.5, suggesting that computer-aided analysis can in some cases be more sensitive to galaxy spirality compared to the human eye. These conclusions are also true for galaxies classified by professional astronomers, as was shown by using the RC3 and NA10 catalogs.

The results also exhibit redshift bias. This bias can be attributed to the quality of the images, as images of nearby galaxies provide higher image quality and therefore manual inspection of these images can be easier compared to images of galaxies with higher redshift, in which the ability of computer-aided analysis to detect subtle differences between gray levels can provide an advantage over the unaided human eye.

\section{Acknowledgments}

Funding for the SDSS and SDSS-II has been provided by the Alfred P. Sloan Foundation, the Participating Institutions, the National Science Foundation, the US Department of Energy, the National Aeronautics and Space Administration, the Japanese Monbukagakusho, the Max Planck Society, and the Higher Education Funding Council for England. The SDSS Web Site is http://www.sdss.org/.

The SDSS is managed by the Astrophysical Research Consortium for the Participating Institutions. The Participating Institutions are the American Museum of Natural History, Astrophysical Institute Potsdam, University of Basel, University of Cambridge, Case Western Reserve University, University of Chicago, Drexel University, Fermilab, the Institute for Advanced Study, the Japan Participation Group, Johns Hopkins University, the Joint Institute for Nuclear Astrophysics, the Kavli Institute for Particle Astrophysics and Cosmology, the Korean Scientist Group, the Chinese Academy of Sciences (LAMOST), Los Alamos National Laboratory, the Max Planck Institute for Astronomy (MPIA), the Max Planck Institute for Astrophysics (MPA), New Mexico State University, Ohio State University, University of Pittsburgh, University of Portsmouth, Princeton University, the United States Naval Observatory and the University of Washington.

\label{last_page}

\end{document}